\documentclass[preprint,onecolumn,nofootinbib]{revtex4}
\pdfoutput=1
\usepackage{float}
\usepackage{graphicx}
\usepackage{amsmath}
\usepackage{amsfonts}
\usepackage{amssymb,ulem}
\usepackage{color}%
\usepackage{dcolumn}

\usepackage{MnSymbol,wasysym}
\usepackage{braket,diagbox}
\usepackage{eurosym}
\usepackage{calrsfs}
\usepackage[usenames,dvipsnames,svgnames]{xcolor}

\newcommand{\RNum}[1]{\uppercase\expandafter{\romannumeral #1\relax}}
\usepackage[colorlinks=true,linkcolor=blue,urlcolor=blue,filecolor=black,citecolor=red,
pdfstartview=FitV,pdftitle={},pdfsubject={},pdfkeywords={},pdfpagemode=None,bookmarksopen=true]{hyperref}

\begin{document}
\baselineskip=0.5 cm
\title{Instability of de-Sitter black hole with massive scalar field coupled to Gauss-Bonnet invariant and the scalarized black holes }

\author{Zhen-Hao Yang}
\email{yangzhenhao$_$yzu@163.com}
\author{Guoyang Fu}
\email{FuguoyangEDU@163.com}
\author{Xiao-Mei Kuang}
\email{xmeikuang@yzu.edu.cn (corresponding author)}
\author{Jian-Pin Wu}
\email{jianpinwu@yzu.edu.cn}
\affiliation{Center for Gravitation and Cosmology, College of Physical Science and Technology,
Yangzhou University, Yangzhou 225009, China}

\begin{abstract}
\vspace*{0.6cm}
\baselineskip=0.5 cm
The black hole scalarization in a special Einstein-scalar-Gauss-Bonnet (EsGB) gravity has been widely investigated in recent years. Especially, the spontaneous scalarization of scalar-free black hole in de-Sitter (dS) spacetime possesses interesting features due to the existence of cosmological horizon. In this work, firstly, we  focus on the massive scalar field perturbation on Schwarzschild dS (SdS) black hole in a special EsGB theory.  By analyzing the fundamental QNM frequency and time evolution of the scalar field perturbation, we figure out the unstable/stable regions in $(\Lambda,\alpha)$-plane as well as in $(m,\alpha)$-plane for various perturbation modes, where $\Lambda$, $\alpha$ and $m$ denote the cosmological constant, the GB coupling strength and the mass of scalar field, respectively.  Then by solving the static perturbation equation, we analyze the bifurcation point at which the SdS black hole supports spherical scalar clouds, and we find that  the bifurcation points match well with $\alpha_c$ on the border of unstable/stable region.
Finally, after addressing that the scalarised solutions could only emerge from the scalar could with node $k\geq 1$. we explicitly construct the scalarized hairy solutions for different scalar masses  and compare the profile of scalar field to the corresponding scalar clouds.
\end{abstract}

\maketitle
\tableofcontents

\section{Introduction}
It is widely accepted that the effects of higher-order curvature terms are significant as we are exploring the strong gravity regime via  detections of gravitational waves and black hole shadows. In theoretical framework, the inclusion of such terms usually involves the well-known ghost problem \cite{stelle}. A counterexample which can be ghost-free is including the Gauss-Bonnet (GB) correction, however, it becomes a topological term in four-dimensional spacetime and has no dynamics when minimally coupled with Einstein-Hilbert action.  One way to make this term contribute to the dynamic in four-dimensional spacetime is to introduce a coupling between the GB term and scalar field \cite{stringT}. The theory which includes this kind of coupling is dubbed Einstein-scalar-Gauss-Bonnet (EsGB) gravity, which has attracted plenty of attention as it admits hairy black holes.  Various black hole solutions and compact objects in four-dimensional EsGB theories were studied in the literatures \cite{Mignemi_1993,Kanti_1996,Torrii_1996,Guo:2008eq,Maeda:2010bj,Ohta:2013lsa,Ayzenberg_2014,Kleihaus_2011,Kleihaus_2016a} and therein.
More recently, the spontaneous scalarization of scalar-free black hole with particular coupling functions in EsGB theory was proposed.
It was addressed that below a certain mass the Schwarzschild black hole background may become unstable in regions of strong curvature, and then a scalarized hairy black hole emerges when the scalar field backreacts to the geometric. The precess evades the well-known no-hair theorems \cite{Antoniou:2017acq,Silva:2017uqg,Doneva:2017bvd}. This proposal on spontaneous scalarization has inspired wide generalizations in the literatures \cite{Herdeiro:2019yjy,Brihaye:2018bgc,Minamitsuji:2018xde,Silva:2018qhn,Andreou:2019ikc,
Minamitsuji:2019iwp,Peng:2020znl,Liu:2020yqa,Doneva:2020qww,Astefanesei:2020qxk,Canate:2020kla,
Hunter:2020wkd,Bakopoulos:2019tvc,Bakopoulos:2020dfg,Lin:2020asf,
Brihaye:2019dck,Guo:2020sdu,Tang:2020sjs,Collodel:2019kkx,Dima:2020yac,Doneva:2020kfv,Herdeiro:2020wei,
Berti:2020kgk,Guo:2020zqm}.

For various reasons, physicists have great interest in de Sitter (dS) spacetime. On one hand, the theoretical model of our expansion universe consists on assuming positive cosmological constant, implying that the physical universe is asymptotically dS. Besides, dS spacetime plays an important role in primordial inflation theory, which is now part of the standard cosmological model \cite{SupernovaCosmologyProject:1998vns,SupernovaSearchTeam:1998fmf}.
On the other hand,  the holographic duality between quantum gravity in dS spacetime
and a conformal field theory on it boundary sheds remarkable application on the
asymptotically dS spacetime \cite{Strominger:2001pn,Witten:2001kn}.

Due to the above consideration, the spontaneous scalarization in EsGB theory has been soon extensively studied in dS spacetime \cite{Bakopoulos:2018nui,Brihaye:2019gla}. It was found that the positive cosmological constant does not change the local conditions for a tachyonic instability of the background black hole, Schwarzschild dS (SdS) in this case, to emerge. In details, in \cite{Bakopoulos:2018nui}, the authors claimed that a regular black hole horizon with a non-trivial hair may be always formed after an analysis in the near-horizon asymptotic regime. But the complete hairy solution was absent, the deep reason of which is not clear. Later in \cite{Brihaye:2019gla},
it was addressed if the scalar field is confined between the black hole and cosmological horizons, then it is not likely to form scalarised black hole solution; while a new hairy black hole was numerically constructed if the scalar field is permitted to extend beyond the cosmological horizon. Thus, it is obvious that the existence of cosmological horizon introduces particular situation in the black hole scalarization in dS spacetime, which deserves further study.

It is noted that though the analysis of (in)stability on the background black hole under the massless scalar field perturbation was present in \cite{Bakopoulos:2018nui,Brihaye:2019gla}, however, the quasi-normal mode (QNM) frequencies and the dynamical evolution of the scalar field in the SdS background is still missing. This paper intends to fill this gap. We shall consider a massive scalar field as a probe field on the SdS black hole and admit the scalar field can extend beyond the cosmological horizon. By computing its  fundamental QNM frequencies and the linear time domain dynamical evolution, we fix the unstable/stable parameter regions in $(\Lambda,\alpha)$-plane and $(m,\alpha)$-plane, respectively for various angular momentum modes. Then we analyze the possible existence of scalar cloud by solving the static perturbation equation. We show that the bifurcation points from the lowest sate of scalar field with node $k=0$ match well with the border of unstable/stable region obtained in the fundamental frequency for various modes massless/massive scalar perturbation. 
Finally, we analyze the possible constraint on the models that can have scalarised solutions and address that it could emerge from the scalar could with node $k\geq 1$. Considering the  backreaction,  we construct the scalarized hairy black hole solution  for different scalar masses from the complete non-linear differential equations, and  we also see how the  scalar cloud evolves into the hairy solution by comparing their profiles.

The remaining of this paper is organized as follows. In section \ref{sec:model}, we briefly present the EsGB model with a positive cosmological constant, and the equations of motion. In section \ref{sec:instability},  we analyze the (in)stability of the SdS black hole under the perturbation of massive scalar field via investigating its fundamental QNM frequencies and the {time evolution}.  We investigate the bifurcation points at which the SdS black hole supports spherical scalar clouds with different modes, and then construct the scalarized hairy black hole solutions for different scalar masses in section \ref{sec:backreaction}. The last section contributes to our conclusion and discussion. We shall work with the units $\hbar=G=c=1$.

\section{Model }\label{sec:model}
The action of EsGB gravity in dS spacetime with the scalar field coupled with the GB invariant is given  by
\begin{equation}
S=\frac{1}{16 \pi }\int d^4x\sqrt{-g}\left[R-2 \Lambda -2\nabla _{\mu }\phi \nabla ^{\mu }\phi -m^2 \phi ^2-f( \phi)  \mathcal{L}_{\text{GB}}\right],
\label{action}
\end{equation}
where $R$ is the Ricci scalar, $\Lambda$ is a positive cosmological constant, $\phi$ is scalar field with mass $m$, $f( \phi)$ is the coupling function, and
\begin{equation}
\begin{split}
\mathcal{L}_{\text{GB}}=R^2-4 R_{\mu \nu } R^{\mu \nu }+R_{\mu \nu \rho \sigma } R^{\mu \nu \rho \sigma }.
\end{split}
\end{equation}
Noted that different forms of  $f(\phi)$ shall give different properties of the EsGB theory. As mentioned in  \cite{Doneva:2017bvd}, to admit Schwarzschild black hole as  background solution, $f(\phi)$ could satisfy the conditions $\frac{df(\phi)}{d\phi}\mid_{\phi=0}=0$ and $\frac{d^2f(\phi)}{d\phi^2}\mid_{\phi=0}=\mathfrak{b}^2>0$, where $\mathfrak{b}$ is a constant. Moreover, one usually assumes that the scalar field vanishes at infinity and normalizes the constant $\mathfrak{b}$ to be  unity. Thus, to fulfill the requirement, we shall follow \cite{Brihaye:2019gla} and choose the simplest form of the coupling function as
\begin{equation}\label{eq-fphi}
f(\phi)=a_{0}-\alpha\phi^2,
\end{equation}
where $\alpha_0$ can be an arbitrary value,  and $\alpha$ is the coupling parameter with dimension $[length]^2$ such that we will use the dimensionless parameter $ \alpha/M^2 \to \alpha$  (Here $M$ is kind of black hole mass with length dimension  as we will see soon.).

Then the equations of motion for the scalar field and the gravity field are
\begin{eqnarray}
\nabla_{\mu}\nabla^{\mu}\phi&=&\frac{1}{2}m^2\phi-\frac{\alpha\phi\mathcal{L}_{\text{GB}}}{2},\label{eq:eom1}\\
R_{\mu\nu}-\frac{1}{2}Rg_{\mu\nu}+\Lambda g_{\mu\nu}&=&2(T_{\mu\nu}^{\phi}+T_{\mu\nu}^{GB})\label{eq:eom2}
\end{eqnarray}
where
\begin{eqnarray}
T_{\mu\nu}^{\phi}&=&\partial_{\mu}\phi\partial_{\nu}\phi-\frac{1}{4}g_{\mu\nu}m^2\phi^2
-\frac{1}{2}g_{\mu\nu}\partial_{\rho}\phi\partial^{\rho}\phi,\\
T_{\mu\nu}^{GB}&=&-2\alpha(-\frac{1}{4}\varepsilon_{\mu\sigma\gamma\tau}R^{\gamma\tau\epsilon\delta}\varepsilon_{\nu\rho\epsilon\delta})\nabla^{\rho}\nabla^{\sigma}f(\phi)
.\end{eqnarray}
In the above sector, we shall  firstly treat the scalar field as perturbation and study the (in)stability of the sector, and then we involve the backreaction of the scalar field to the geometry and construct the scalarized hairy black hole solution.

\section{Analysis on the (in)stability from massive scalar field perturbation }\label{sec:instability}

With $\phi=0$, the above equations admit the Schwarzschild de-Sitter (SdS) black hole solution
\begin{eqnarray}
ds^2&&= -g(r) d t^2+ \frac{ d r^2}{g(r)}+r^2 \left(  \text{sin}^2\theta d\theta^2 +d \phi ^2\right)\\
&&\mathrm{with} ~~~~g(r)= 1-\frac{2 M}{r}-\frac{\Lambda  r^2}{3}
\end{eqnarray}
where the constant $M$ is the black hole mass. In this background, the GB term is evaluated as
\begin{equation}
\mathcal{L}_{\text{GB}}=\frac{48 M^2}{r^6}+\frac{8 \Lambda ^2}{3}.
\end{equation}
Depending on the model parameters, $g(r)=0$ could have two positive real roots, $(r_e,r_c)$,  and one negative real root, $r_o$. The positive roots, $r_e$ and $r_c$, represent the black hole event horizon and cosmological horizon, respectively, of the SdS black hole. Note that these two horizons only hold when $3M\sqrt{\Lambda}<1$, known as the Nariai limit, within which their analytic formulas are
\begin{equation}
	r_e=\frac{2}{\sqrt{\Lambda}}\cos\left(\frac{1}{3} \cos^{-1}\left(3M\sqrt{\Lambda}\right)+\frac{\pi}{3}\right)\;, \quad r_c=\frac{2}{\sqrt{\Lambda}}\cos\left(\frac{1}{3} \cos^{-1}\left(3M\sqrt{\Lambda}\right)-\frac{\pi}{3}\right).
\end{equation}
Approaching the Nariai limit, the event horizon and the cosmological horizon tend to merge into one horizon; and beyond the limit, no black hole horizon exists.

Considering $r_c>r_e>r_o$, the metric function $g(r)$ can be further expressed as,
\begin{equation}
g(r)=\frac{\Lambda}{3 r}(r-r_e)(r_c-r)(r-r_o).
\end{equation}
Introducing the surface gravity $\kappa_i =\frac{1}{2}\mid g'(r) \mid_{r=r_i}$ associated with each root $r=r_i~(i=e,c,o)$, one could calculate the tortoise coordinate $r_{*}= \int g^{-1}(r) \, dr$ in an analytic form
\begin{equation}
r_{*}(r)= \frac{1}{2\kappa_e} \ln\left(\frac{r}{r_e}-1\right)-\frac{1}{2\kappa_c} \ln\left(1-\frac{r}{r_c}\right)+\frac{1}{2\kappa_o} \ln\left(\frac{r}{r_o}-1\right)
\end{equation}
which will play an important role in our later numerical study.

Then we consider a small scalar field perturbation on the background of SdS black hole in the linear regime, which is governed by the  covariant equation
\begin{equation}\label{eq-14}
\square \phi =\frac{\partial f(\phi )}{\partial \phi } \frac{\mathcal{L}_{\text{GB}}}{4}+\frac{m^2}{2}\phi \equiv \mu _{\text{eff}}^2 \, \phi.
\end{equation}
Here the box denotes the d'Alembertian operator and the effective mass is
\begin{equation}\label{eq-meff1}
\mu _{\text{eff}}^2= \frac{m^2}{2}-\frac{ \alpha}{2}  \left(\frac{8 \Lambda ^2}{3}+\frac{48 M^2}{r^6}\right).
\end{equation}
The tachyonic instability may occur only when $\mu _{\text{eff}}^2<0$, which requires $\alpha>0$. As addressed in \cite{Brihaye:2019gla}, this instability may trigger the emergence of a hairy solution on the SdS background via the spontaneous scalarization process, though the authors only considered the massless $r-$dependent scalar field. The possible parameter regions which give unstable situation and how the scalar field grows up were not present.  Here we shall answer those questions by studying the dynamical perturbation of scalar field with various modes in both frequency domain and time domain.


\subsection{Preliminary preparation}\label{sec:numerical method}
\subsubsection{Frequency domain analysis}\label{sec:W-Analysis}
To analyze the (in)stability in frequency domain, one usually decompose the scalar field as
\begin{equation}\label{eq-decomposation0}
\phi(t,r,\theta,\psi)=\sum_{\ell\mathfrak{m}}\frac{R_{\ell\mathfrak{m}} (r)}{r} Y_{\ell\mathfrak{m}} (\theta,\psi)e^{-i\omega t}
\end{equation}
where $Y_{\ell\mathfrak{m}} (\theta,\psi)$ is the spherical harmonic function with angular momentum $l$ and azimuthal number $\mathfrak{m}$.
Working under the tortoise coordinate $r_{*}$, we obtain that each wave function $R$ satisfies
\begin{equation}\label{eq-eom1}
\frac{\partial^2 R(r)}{\partial r_{*}^2}+\left(\omega^2-V_{eff}(r)\right)R(r)=0,
\end{equation}
where the effective potential is
\begin{equation}\label{eq-V1}
V_{eff}(r)=g(r)\left[\frac{\ell(\ell+1)}{r^2}+\frac{g^{'}(r)}{r}+\mu _{\text{eff}}^2  \right],
\end{equation}
and it is not dependent of the azimuthal number $\mathfrak{m}$ because of the spherical symmetry. This potential could  present a negative well between $r_e$ and $r_c$, which plays an important role in the instability of the SdS black hole under this perturbation as we will see soon. The asymptotic behavior of the perturbation near the horizons are
\begin{equation}
R(r\to r_e)\sim e^{-i\omega r_{*} }~~~~\mathrm{and}~~~~R(r\to r_c)\sim e^{i\omega r_{*} }
\end{equation}
which correspond to ingoing and outgoing boundary condition near the event horizon and cosmological horizon, respectively. It is known that only discrete eigenfrequencies $\omega=\omega_R+i \omega_I$ (QNM frequency), where $\omega_R$ and $\omega_I$ respectively denote the real part and imaginary part of the QNM frequency,  satisfy the perturbation equation and the boundary conditions. Once $\omega_I>0$, the amplitude of the perturbation will grow up, implying that the black hole is unstable under this perturbation. There are many methods developed to compute the QNM frequencies, for instance, WKB method, shooting method, Horowtiz-Hubeny method, AIM method, spectral method, etc., and readers could refer to \cite{Berti:2009kk,Konoplya:2011qq} and therein for nice reviews. In this work, we will employ the spectral method, which has been well described in \cite{Jansen:2017oag}. Moreover, we shall also testify our results with the time domain evolution.

\subsubsection{Time domain analysis}\label{sec:T-Analysis}

To study the dynamical evolution of the perturbed scalar field, we decompose the scalar field  as
\begin{equation}\label{eq-decomposation}
\phi(t,r,\theta,\psi)=\sum_{\ell \mathfrak{m}}\frac{R_{\ell\mathfrak{m}} (t,r)}{r} Y_{\ell\mathfrak{m}} (\theta,\psi).
\end{equation}
In the tortoise coordinate, the perturbation equation reduces to
\begin{equation}\label{eq-eom2}
\left(-\frac{\partial^2}{\partial t^2}+\frac{\partial^2}{\partial r_{*}^2}-V_{eff}(r)\right)R(t,r)=0
\end{equation}
where $V_{eff}(r)$ has been defined in \eqref{eq-V1}.

We have to solve the wave equation numerically since there is no analytic form of this time-dependent wave equation. We adopt the discretization method proposed in \cite{Zhu:2014sya}, and discretize the wave equation \eqref{eq-eom2} by defining $R(r_*,t)=R(j\Delta r_*,i\Delta t)=R_{j,i}$ , $V\left(r(r_*)\right)=V(j\Delta r_*)=V_j$. Then it can be deformed as
\begin{equation}
-\frac{(R_{j,i+1}-2R_{j,i}+R_{j,i-1})}{\Delta t^2}+\frac{(R_{j+1,i}-2R_{j,i}+R_{j-1,i})}{\Delta r_*}-V_j R_{j,i}+O(\Delta t^2)+O(\Delta r^2_*)=0.
\end{equation}
With the initial Gaussian distribution $R(r_*,t=0)=\exp[-\frac{(r_*-a)^2}{2b^2}]$ and $R(r_*,t<0)=0$, we can derive the evolution of $R$ by
\begin{equation}\label{eq-evolution1}
R_{j,i+1}=-R_{j,i-1}+\frac{\Delta t^2}{\Delta r^2_*} R_{j+1,i}+R_{j-1,i}+\left(2-2\frac{\Delta t^2}{\Delta r^2_*}-\Delta t^2 V_j\right) R_{j,i}.
\end{equation}
For the sake of the numerical precision, we shall fix $\frac{\Delta t^2}{\Delta r^2_*}=0.5$ to fit the von
Neumann stability conditions, and set $a=10,b=3$ in the profile of Guassian wave.

\subsection{The QNM frequencies and instability}\label{sec:results}
We will study how the coupling parameter and the mass of scalar field affect the (in)stability of SdS black hole. We will also consider different cosmological constants and angular momentums of the perturbation modes.  To this end, we fix $M=1$ without loss of generality.
\subsubsection{$\ell-$dependence}
In this subsection, we compute the \emph{fundamental (with overtone $n=1$)} QNM frequency of different $\ell-$modes perturbation around the SdS black hole. We focus on the effect of $\ell$ on the GB coupling dependent QNM frequencies by fixing $m=0$ and  $\Lambda=0.1$.

\begin{figure}[H]
 \centering
  \includegraphics[width=5cm]{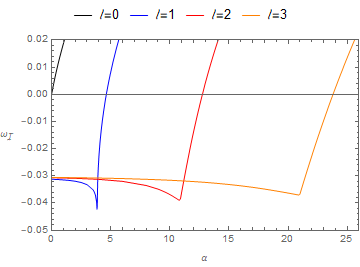}\hspace{1cm }
  \includegraphics[width=5cm]{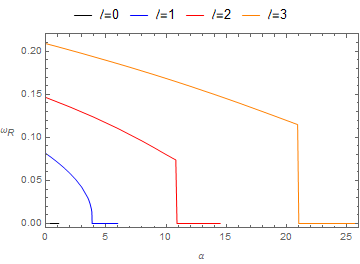}
	  \caption{The fundamental QNM frequency as a function of the GB coupling for different $\ell$ modes.}
 \label{fig:a-w-changl}
\end{figure}

The results are shown in FIG. \ref{fig:a-w-changl}.  For the $\ell = 0$ mode, both $\omega_R$ and $\omega_I$ are zero in minimal coupling case implying that the
mode is prone to instability. As the coupling parameter, $\alpha$, increases, the black lines shows that $\omega_I$ becomes positive while $\omega_R$ is still zero. This phenomena means that the $\ell = 0$ mode is always unstable. For the $\ell \neq 0$ modes,  the left plot shows that as $\alpha$ increases, $\omega_I$ first decreases and then increases. There exists a critical value, $\alpha_c$, at which $\omega_I=0$ for each mode. When $\alpha>\alpha_c$,  $\omega_I$  becomes positive implying that the corresponding perturbation could grow up to destabilize the background SdS black hole. $\alpha_c$ is larger for modes with larger $\ell$ and the samples are listed in table \ref{table:l-ac}, which implies that under larger $\ell$ mode perturbation, stronger GB coupling is called for to trigger the instability.

Moreover, comparing the two plots in FIG. \ref{fig:a-w-changl}, we see that the cases with decreasing $\omega_I$ have non-vanishing  $\omega_R$, while the cases with increasing $\omega_I$ possess purely imaginal QNM frequency. As addressed in \cite{Aragon:2020tvq}, the former cases are dubbed photon sphere modes while the latter are dS modes.
In our study, since the critical $\alpha$ from photon sphere mode to dS mode is always smaller than $\alpha_c$, so the unstable perturbation modes are all dS modes, which is an interesting phenomena deserving further study.

\begin{table}[H]
\center{
\begin{tabular}{|c|c|c|c|c|c|c|c|c|}
  \hline
  $\ell$ & $0$ & $1$ & $2$ & $3$ & $4$ & $5$ & $6$  \\  \hline
 $\alpha_c$  & $-$ & $4.63$ &$12.75$ &$23.78$ & $37.84$& $54.93$ & $75.09$  \\
  \hline
\end{tabular}
\caption{The critical coupling for different momentum angular.}\label{table:l-ac}}
\end{table}

\begin{figure}[H]
 \centering
  \includegraphics[width=5cm]{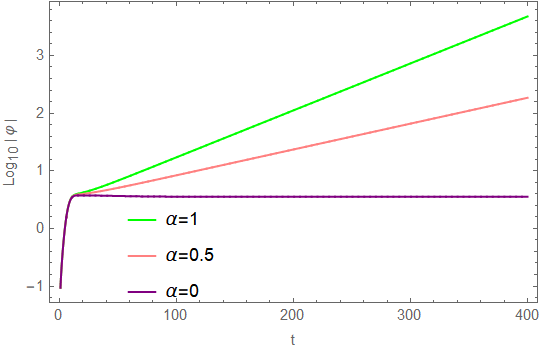}\hspace{1cm}
  \includegraphics[width=5cm]{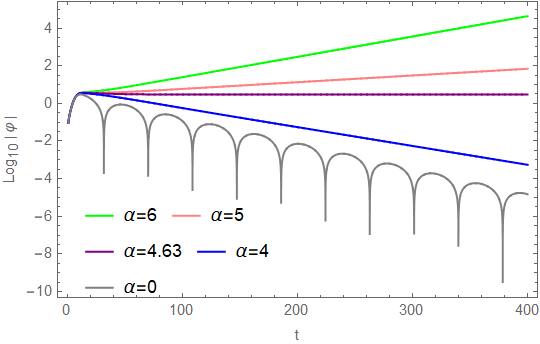}\\
  \includegraphics[width=5cm]{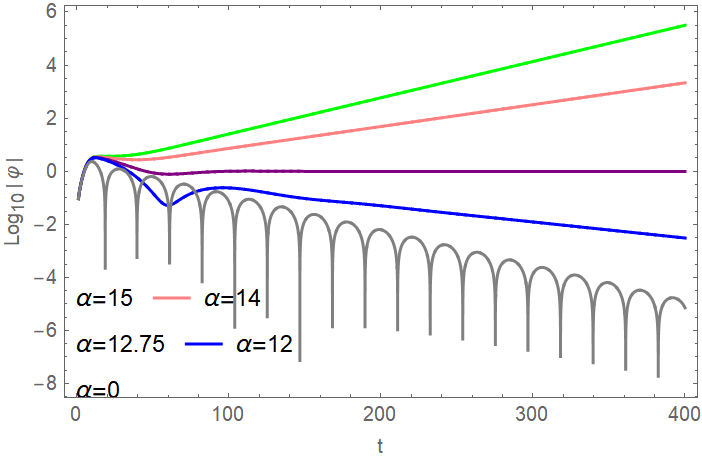}\hspace{1cm}
  \includegraphics[width=5cm]{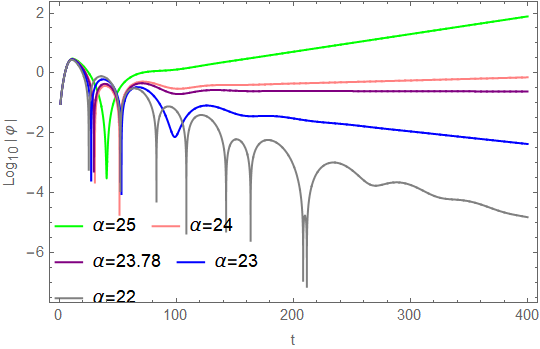}
	  \caption{The time evolution of the perturbation modes with $\ell=1$ , $\ell=2$ and  $\ell=3$  from left to right.}
 \label{fig:t-phi-m0A01}
\end{figure}

We also directly calculate the time evolution of the perturbation field and further reveal the instability of SdS in EsGB gravity. The pedagogy on evolutionary analysis has been shown in subsection \ref{sec:T-Analysis}. The evolutions of the perturbation with different $\ell$ in log plot are shown in FIG. \ref{fig:t-phi-m0A01}.  For $\ell=0$ mode, non-vanishing GB coupling makes  the perturbation grow as time, meaning that the system is unstable. For $\ell\neq 0$, when $\alpha<\alpha_c$, the perturbation will decay as the time evolves, but it grows up when  $\alpha>\alpha_c$. The phenomenon indicates that the system will become unstable under the perturbed modes once the GB coupling is larger than the corresponding $\alpha_c$. These findings in time domain are consistent with those in frequency domain analysis.
The effect of the GB coupling on the dynamical evolution is different from that of the non-minimally coupled to curvature studied in \cite{Brady:1999wd} which always makes the evolution decay in SdS black hole.

\begin{figure}[ht!]
 \centering
  \includegraphics[width=5cm]{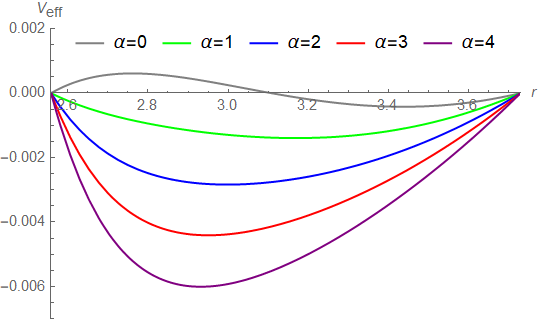}\hspace{0.5cm}
  \includegraphics[width=5cm]{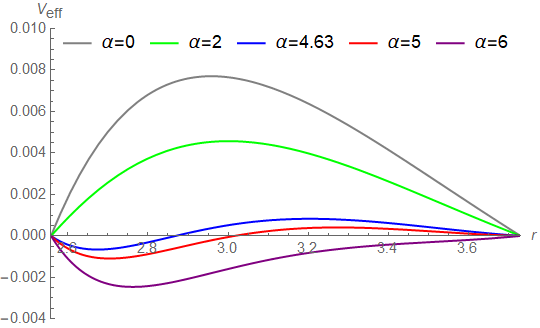}\hspace{0.5cm}
  \includegraphics[width=5cm]{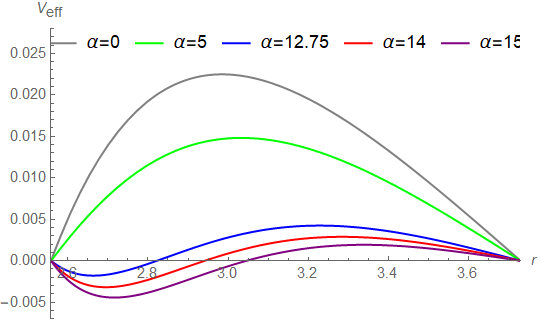}
	  \caption{The profile of the effective potential between the event horizon and cosmological horizon for $\ell=0,1,2$ from left to right.}
 \label{fig:rVm0}
\end{figure}

The effect of the GB coupling on the (in)stability of the scalar perturbation could be explained by analyzing the effective potential \eqref{eq-V1}. It is straightforward to obtain that  the effective potential with a fixed radius near the event horizon could be suppressed by larger $\alpha$ but enhanced by larger $\ell$. Then in FIG. \ref{fig:rVm0}, we explicitly show the profile of the potential between the event horizon and cosmological horizon. It is obvious that in each case, when $\alpha$ is smaller than a certain value, the effective potential  is always positive. As $\alpha$ increases, a negative potential well would form, and becomes more deeper as $\alpha$ is further enlarged. It is noted that comparing $\alpha_c$ in the frequency as well as time domain analysis, the critical value of $\alpha$ for the emergence of negative potential well in FIG. \ref{fig:rVm0} is smaller in all cases. This is because  the negative potential well is not a sufficient condition for the instability. Only deep enough potential well could  help the scalar to collect near the event horizon and finally trigger the instability of system.

\subsubsection{$\Lambda-$dependence}
\begin{figure}[H]
 \centering
  \includegraphics[width=5.cm]{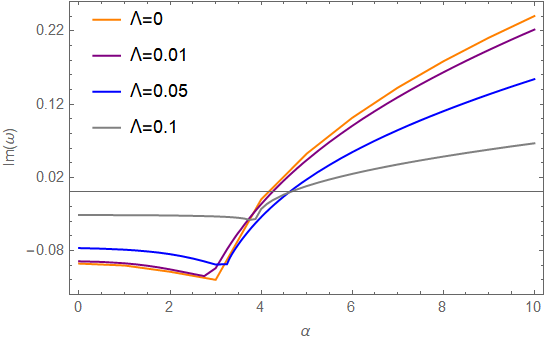}\hspace{0.5cm }
  \includegraphics[width=5.cm]{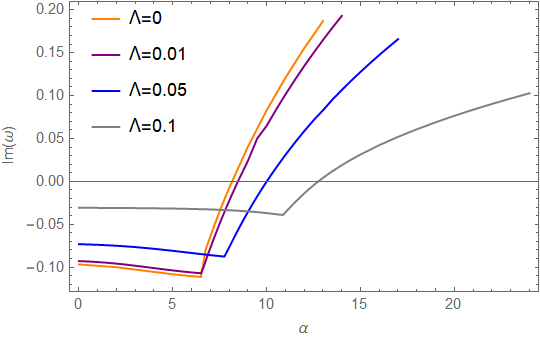}\hspace{0.5cm }
  \includegraphics[width=5.cm]{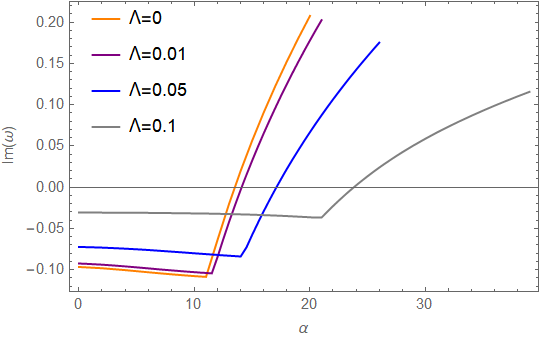}
	  \caption{The fundamental QNM frequency as a function of GB coupling for different cosmological constant. The modes from left to right are $\ell=1,2$ and $3$, respectively. }
 \label{fig:alpha-wI-Lambda}
\end{figure}

\begin{figure}[H]
	\centering
	\includegraphics[width=5.cm]{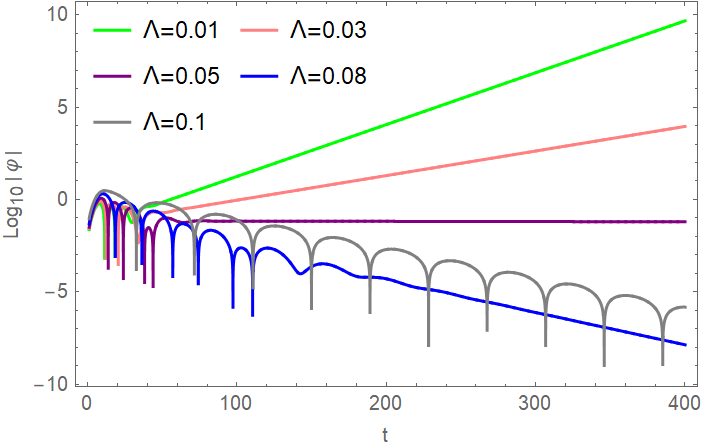}\hspace{0.5cm }
	\caption{The time evolution of perturbation mode with $\ell=2$. Here the GB coupling parameters is  fixed as $\alpha=10$. }
	\label{fig:tphim0L2alpha10}
\end{figure}

We then study the effect of the cosmological constant $\Lambda$ on the fundamental QNM frequency. The results are shown in FIG. \ref{fig:alpha-wI-Lambda}, which shows that the effect of $\Lambda$ on various $\ell$-modes is similar. Specifically, for small $\alpha$, $\omega_I$ is negative and larger for larger $\Lambda$. It means that the perturbation in the spacetime with larger $\Lambda$ can live longer and then decays, since the lifetime is connected with the QNM frequency via $\tau\sim1/|\omega_I|$. As $\alpha$ increases,  $\omega_I$ increases and there exists an intersection for different $\Lambda$. As $\Lambda$ increases, the critical value $\alpha_c$ at which  $\omega_I$ transits from negative to positive increases. Then for large enough $\alpha$, $\omega_I$ become positive and is smaller for larger $\Lambda$. This indicates that in SdS black hole with larger $\Lambda$, the perturbation has longer relax time and then grows up. The above picture can also explicitly reflected in the time domain analysis, see FIG. \ref{fig:tphim0L2alpha10} for the $\ell=2$ mode with $\alpha=10$ as an example.

\begin{figure}[H]
 \centering
  \includegraphics[width=6.cm]{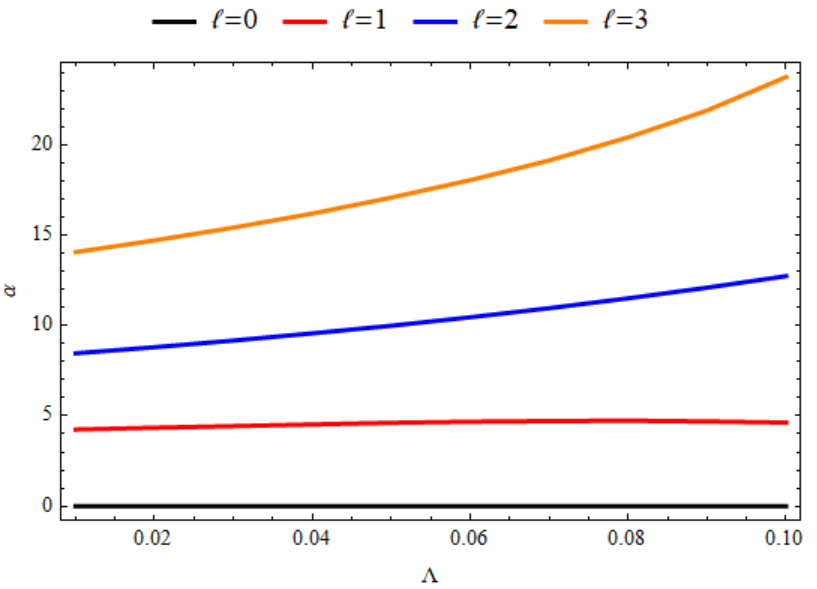}
	  \caption{The unstable/stable region of SdS black hole under massless scalar perturbation with $\ell=0,1,2$ and $3$ from fundamental QNM frequency.  }
 \label{fig:lambda-alpha}
\end{figure}

It is worthwhile to point out that for $\alpha=0$, i.e., in Einstein gravity, our QNM frequencies with different $\ell$ and $\Lambda$ calculated via spectral method match well with the results computed by WKB method and Prony method in \cite{Abdalla:2003db,Zhidenko:2003wq}. Then in EsGB theory with $\alpha\neq0$, by scanning the cosmological constant inside the Nariai limit,  we extract $\alpha_c$ and figure out the unstable/stable region in $\Lambda-\alpha$ plane under massless scalar perturbation. The results for samples of $\ell$ are shown in FIG. \ref{fig:lambda-alpha}. Under the perturbation mode with each $\ell$, the instability can be triggered in the region above each line (the shading region), below which the system is stable. It is obvious that for the modes with larger $\ell$, the system could be stable in a wider parameters region in $\Lambda-\alpha$ plane.
\subsubsection{$m-$dependence}
\begin{figure}[H]
	\centering
	\includegraphics[width=5.cm]{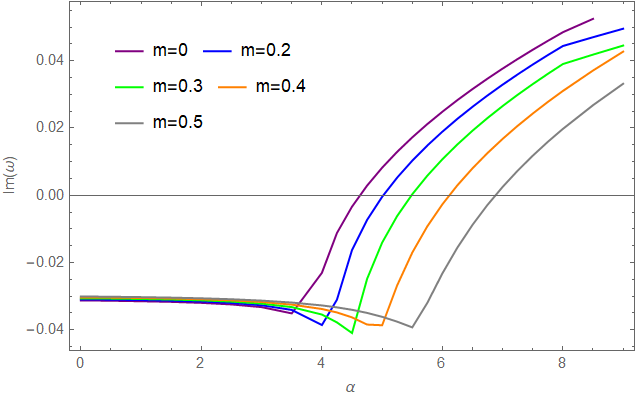}\hspace{0.5cm }
	\caption{The fundamental QNM frequency for $\ell=1$ mode as a function of GB coupling for different mass of scalar perturbation.  }
	\label{fig:alpha-w-m}
\end{figure}

In this subsection, with fixed $\Lambda=0.1$, we shall turn on the mass of the scalar field and study its effect. The fundamental QNM frequencies for $\ell=1$ mode with different $m$ is shown in FIG. \ref{fig:alpha-w-m}.
It shows that the rule is similar to that for massless case, namely, as the GB coupling increases, $\omega_I$ first decreases and then increases. And for larger $m$, the turning value of $\alpha$ from decreasing to increasing is larger, so is $\alpha_c$ at which $\omega_I$ crosses the horizonal axis. This indicates that for the scalar field with larger mass, stronger GB coupling is required to destabilize the SdS black hole.

\begin{figure}[H]
	\centering
	\includegraphics[width=5.cm]{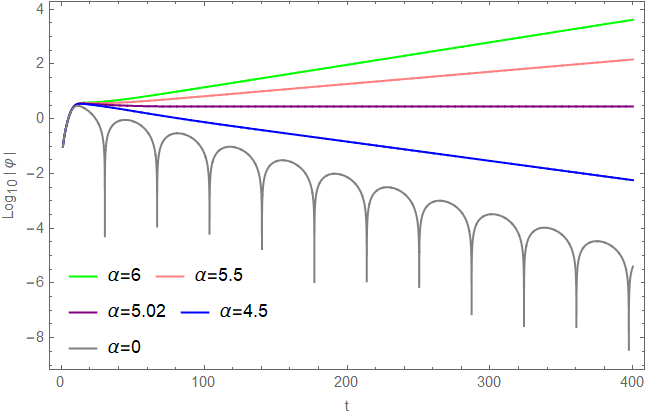}\hspace{1cm }
	\includegraphics[width=5.cm]{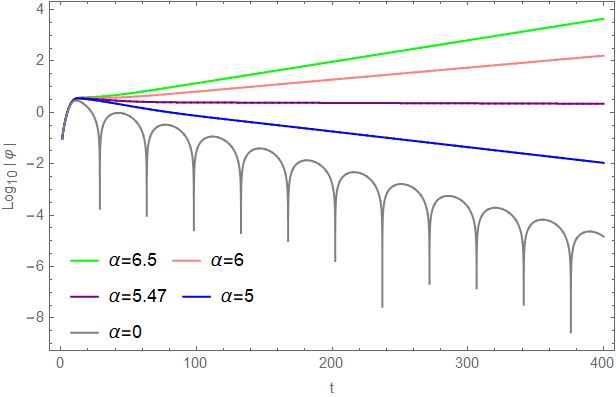}\\
	\includegraphics[width=5.cm]{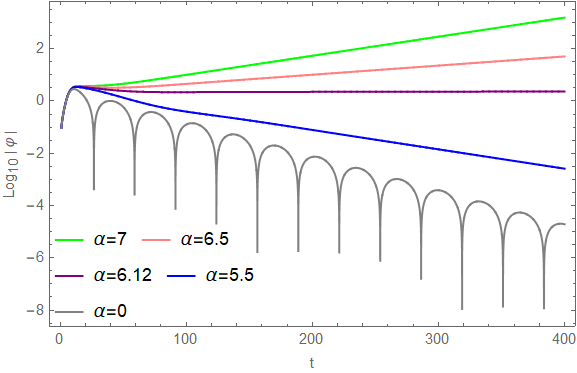}\hspace{1cm }
	\includegraphics[width=5.cm]{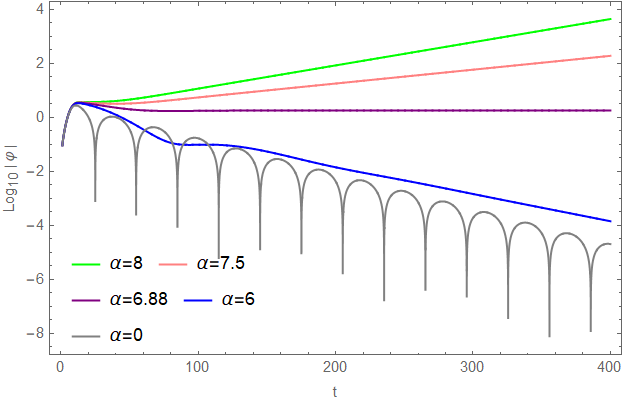}
	\caption{The dynamical evolution of massive scalar field in log plot. The upper-left, upper-right, lower-left and lower-right plots correspond to $m=0.2,0.3,0.4$ and $0.5$ respectively. }
	\label{fig:t-phi-A01L1}
\end{figure}

\begin{figure}[H]
	\centering
	\includegraphics[width=5.cm]{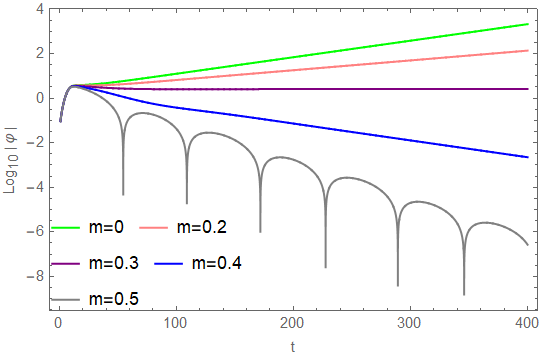}\hspace{0.5cm }
	\caption{The time evolution of the $\ell=1$ mode with fixed $\alpha=5.49$. }
	\label{fig:tphiA01L2alpha5}
\end{figure}

This conclusion is also testfied by the dynamical evolution of the massive scalar field in FIG. \ref{fig:t-phi-A01L1}. Each plot shows that when $\alpha$ is smaller than a certain $\alpha_c$, the scalar field would decay as the time evolves; while the scalar field will grow up when $\alpha$ is larger than $\alpha_c$. The growth of scalar field could finally destabilize the SdS black hole and trigger spontaneous scalarization. Comparing the value of $\alpha_c$ in each plot, it is obvious larger $m$ corresponds to larger $\alpha_c$ which matches the findings in FIG. \ref{fig:alpha-w-m}. More explicit effect of $m$ on the dynamical evolution with fixed $\alpha$ is shown in FIG. \ref{fig:tphiA01L2alpha5}. This mass dependence is reasonable because even in Einstein gravity, increasing
$m$ would enhance $\omega_I$ which is always negative and decrease the scalar field damping rate, such that the corresponding modes live longer \cite{Toshmatov:2017qrq}.


\begin{figure}[H]
	\centering
	\includegraphics[width=6.cm]{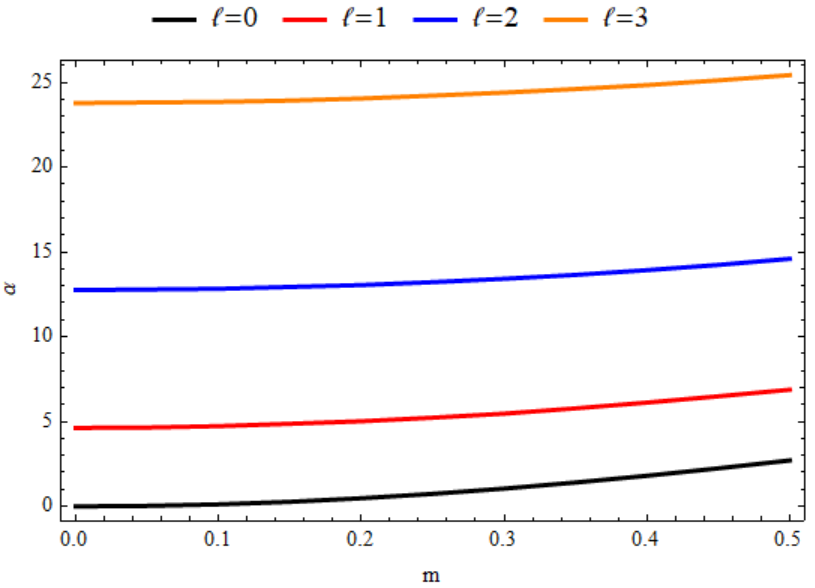}
	\caption{The unstable/stable region of SdS black hole under massive scalar perturbation with $\ell=0, 1,2$ and $3$ from fundamental QNM frequency.  }
	\label{fig:m-alpha}
\end{figure}

For other modes , similar phenomena could be seen as for $\ell=1$ mode. So instead of repeating the analysis, we collect $\alpha_c$ with different $m$ for $\ell=0,1,2$ and $3$ modes, and then draw the unstable/stable region in $m-\alpha$ plane in FIG. \ref{fig:m-alpha}. For each mode, the SdS black hole could be unstable when the parameters are in the regime above each line.

\section{Scalarized hairy black hole solution}\label{sec:backreaction}

In this section, we shall fix the existence of scalar clouds and the scalarized black hole, whose  profile functions  between the black hole horizon and cosmological horizon will be explicitly  constructed.

\subsection{Bifurcation points and scalar clouds}\label{sec:cloud}
The tachyonic instability provides the possibility of the scalar field to be non-trivial
between the black hole and the cosmological horizon. Then we shall go on to seek the bound states (scalar clouds) of the test scalar around the SdS black hole.  To obtain the scalar clouds with various modes, we only need to solve the perturbation equation  \eqref{eq-eom1} in static case with appropriate boundary conditions.  With the two horizons $r_e$ and $r_c$,  the equation is rewritten as
\begin{equation}\label{eq-cloudEQ}
\phi''(r)+\frac{r_c \left(r_e-2 r\right)(r_c+ r_e)-2 r r_e^2+4 r^3}{r \left(r-r_c\right) \left(r-r_e\right) \left(r_c+r_e+r\right)}\phi'(r)+
\frac{\left(r_c r_e+r_c^2+r_e^2\right) \left(\ell^2+\ell+r^2\mu _{eff}\right)}{r \left(r-r_c\right) \left(r-r_e\right) \left(r_c+r_e+r\right)}\phi'(r)=0
\end{equation}
where we recover $\phi(r)=R(r)/r$ and the effective mass is
\begin{equation}
\mu_{eff}=\frac{m^2}{2}-\frac{6 \alpha  \left(r_c^2 r_e^2 \left(r_c+r_e\right){}^2+2 r^6\right)}{r^6 \left(r_c r_e+r_c^2+r_e^2\right){}^2}.
\end{equation}
The regularity  near the black hole horizons give us
\begin{eqnarray}
&&\phi(r\to r_e)=\phi_{h}+\tilde{\phi}_{h}(r-r_e)+\mathcal{O}(r-r_e),\\
&&\phi(r\to r_c)=\phi_{c}+\tilde{\phi}_{c}(r-r_c)+\mathcal{O}(r-r_c)
\end{eqnarray}
where $\tilde{\phi}_{h} (\tilde{\phi}_{c1})$ can be solved to be dependent on $\phi_{h} (\phi_{c})$ .

\begin{figure}[H]
 \centering
  \includegraphics[width=5cm]{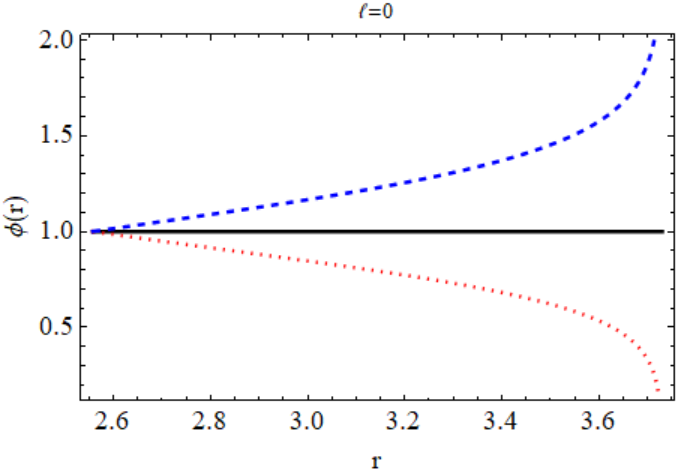}\hspace{0.5cm }
    \includegraphics[width=5cm]{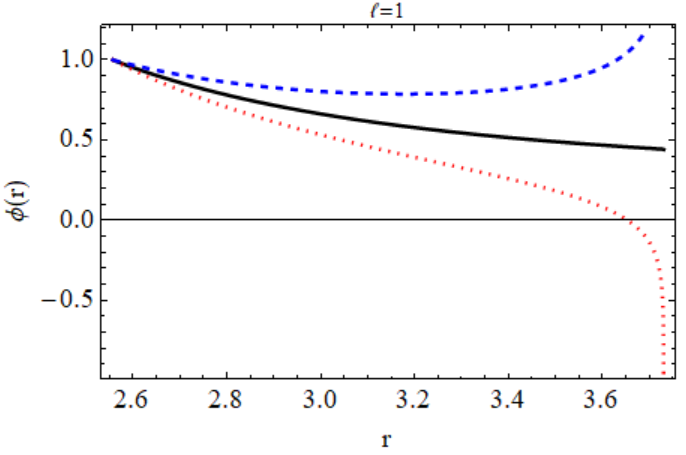}
	  \caption{The scalar field profile of the  lowest state for $\alpha=  bifurcation~point~\alpha_c$ (black), $a=\alpha_c+0.5$ (red) and $a=\alpha_c-0.5$ (blue) with $m=0$ , $\Lambda=0.1$. The left panel is for  $\ell=0$
with $\alpha_c=0$ while  the right panel is for  $\ell=1$ with $\alpha_c=4.63$.  Here we have set $M=1$ and $\phi(r_e)=1$.}
 \label{fig:scalarcloud}
\end{figure}

It is known that in the construction of scalar cloud, solving \eqref{eq-cloudEQ} is in essence  an eigenvalue problem \cite{Silva:2017uqg, Cunha:2019dwb}.  For given $\ell$, $m$ and $\Lambda$, by imposing smoothness for the scalar filed at the  horizons, one can select  a discrete set of  $\alpha/M^2$, which give  a discrete set scalar field profile characterized by different nodes $k$ in between the two horizons. Those selected  $\alpha/M^2$ could be known as  bifurcation points at which SdS black hole supports a spherical scalar could.
In order to give an explicit picture, in FIG.\ref{fig:scalarcloud} we show  typical profiles of  scalar field  with the selected $\alpha/M^2$ for the lowest state with $k=0$ and two other $\alpha/M^2$ with $m=0$, $\Lambda=0.1$ and $\ell=0,1$. It is obvious that only the selected $\alpha/M^2$ corresponds to a smooth scalar profile between the event horizon and cosmological horizon, giving the scalar clouds which satisfy the perturbed equation and the imposed boundary condition.

\begin{figure}[H]
 \centering
  \includegraphics[width=5cm]{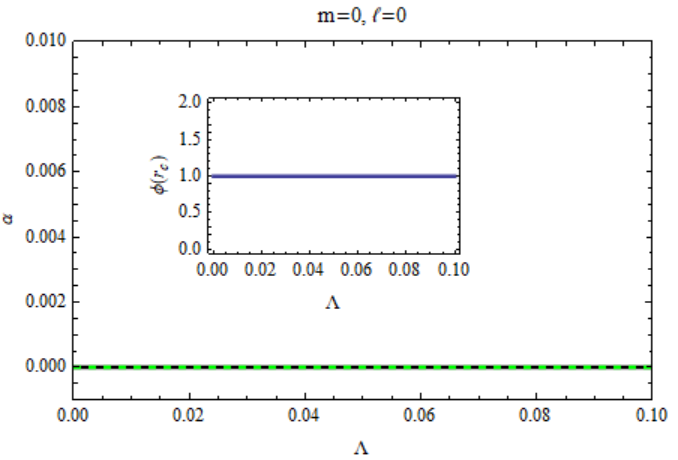}\hspace{0.5cm }
  \includegraphics[width=5cm]{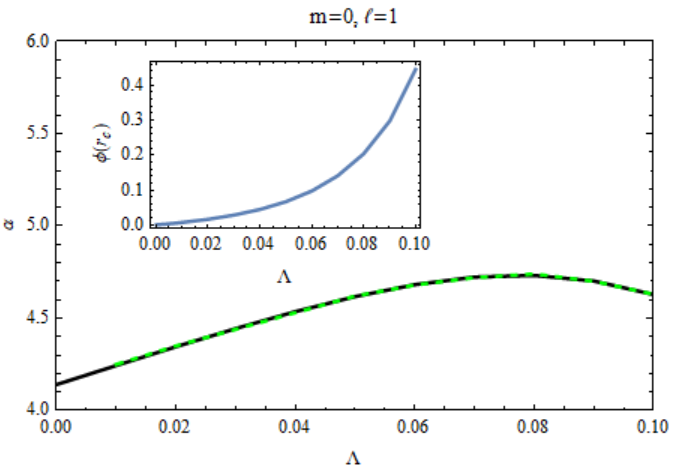}\\
  \includegraphics[width=5cm]{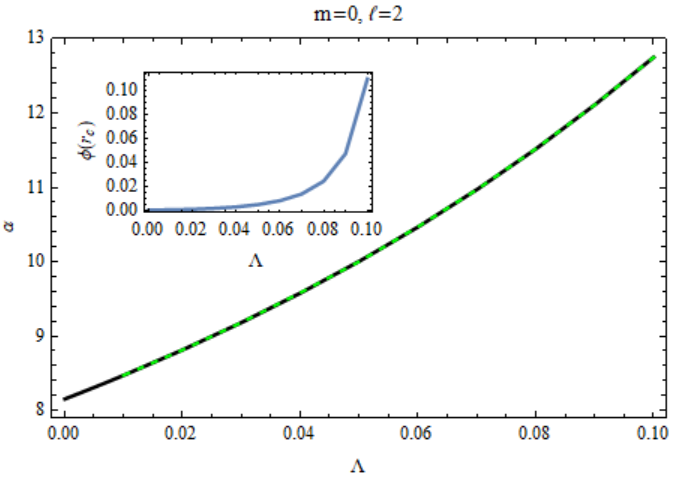}\hspace{0.5cm }
  \includegraphics[width=5cm]{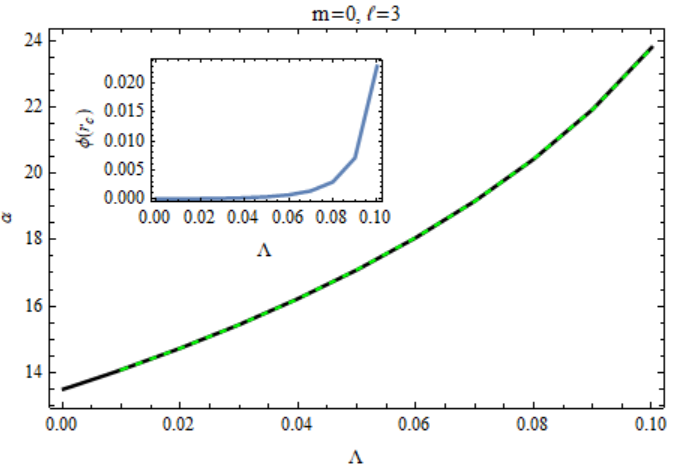}
	  \caption{The coupling parameter against the cosmological constant for the critical SdS black hole that admits a spherical massless scalar could with the lowest state with $k=0$ for various modes, and the insets exhibit the corresponding values of scalar field at $r_c$. Here we have set $M=1$ and $\phi(r_e)=1$.}
 \label{fig:cloud-m0}
\end{figure}

\begin{figure}[H]
 \centering
  \includegraphics[width=5cm]{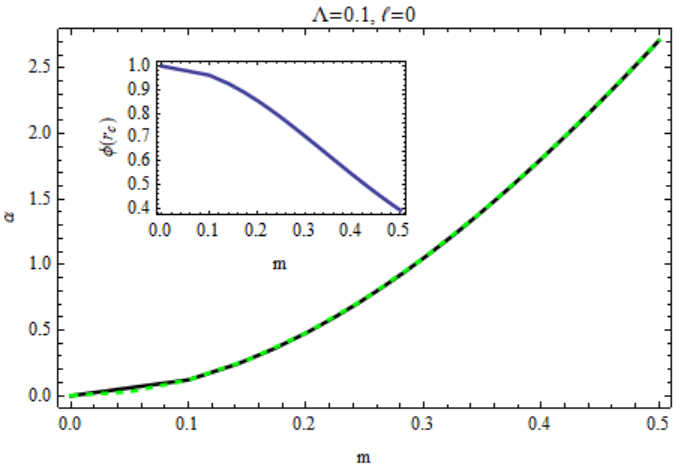}\hspace{0.5cm }
  \includegraphics[width=5cm]{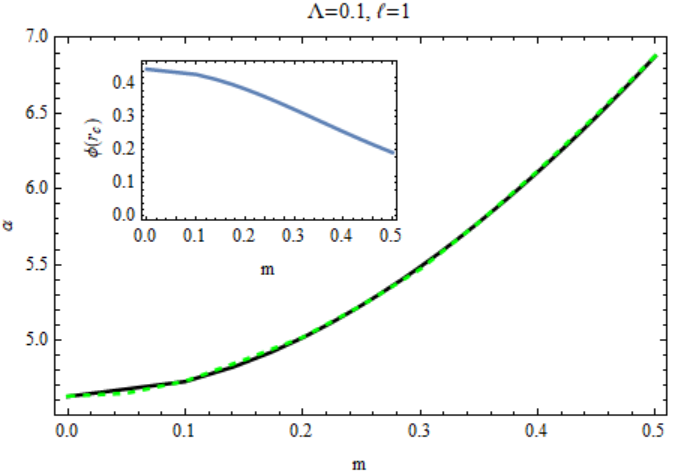}\\
  \includegraphics[width=5cm]{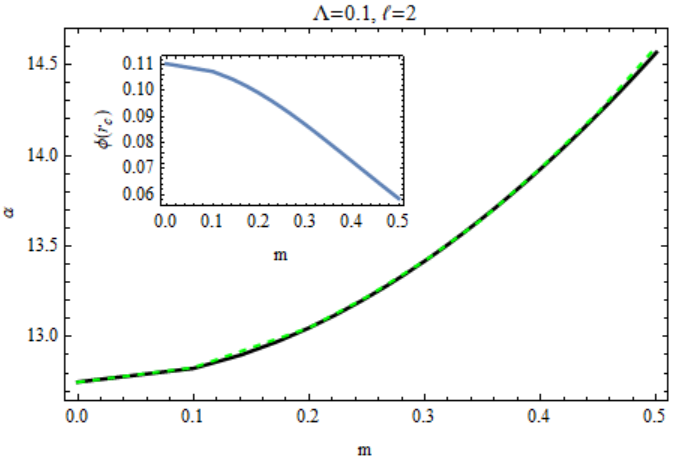}\hspace{0.5cm }
  \includegraphics[width=5cm]{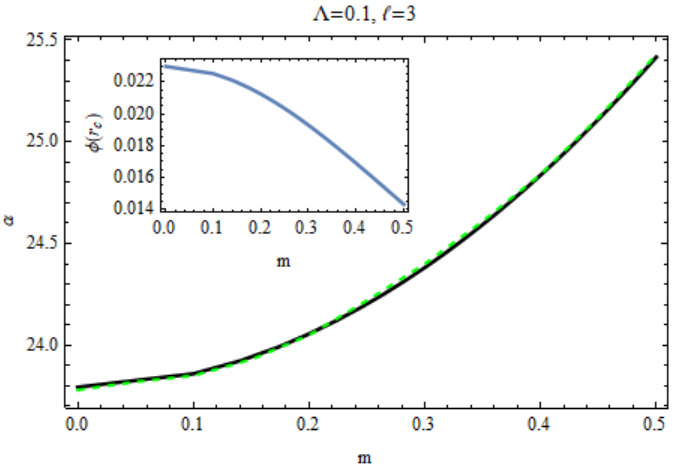}
	  \caption{The coupling parameter against the mass of scalar field for the critical SdS black hole that admits a spherical massive scalar could with the lowest state with $k=0$  for various modes, and the insets exhibit the corresponding values of scalar field at $r_c$. Here we have set $M=1$ and $\phi(r_e)=1$.}
 \label{fig:cloud-A01}
\end{figure}

To study the bifurcation points for various modes and masses of scalar field, we focus on the lowest state with $k=0$ for all cases.
We show the results for $\ell=0,1,2,3$ modes in FIG. \ref{fig:cloud-m0}. In each panel, the green dashed curve denotes the border of unstable/stable region we obtained in the analysis from fundamental frequencies (see the corresponding curves in FIG.\ref{fig:lambda-alpha}). In FIG.\ref{fig:cloud-A01}, we exhibits the effect of scalar  mass on the bifurcation point and the scalar field value at the cosmological horizon for scalar clouds, and again the green dashed curve in each plot denotes the border of unstable/stable region obtained from  fundamental frequencies (see the corresponding curves in FIG.\ref{fig:m-alpha}). As we expect, the bifurcation points for various mode scalar clouds match well with the critical value $\alpha_c$ for the unstable/stable region obtained from the fundamental frequency and time domain.

\subsection{ Construction of scalarized hairy black hole}
The existence of instability and smooth configurations cannot really guarantee the existence of the scalarized hairy solutions, but the authors of \cite{Brihaye:2019gla} gave possible constraints on this model in the case with massless zero mode scalar field. As aforementioned, the requirement of effective mass $\mu^2_{\text{eff}}<0$ should be fulfilled for tachyonic instability. Morevover, following \cite{Brihaye:2019gla}, we integrate \eqref{eq-14} along a hypersurface $V$ bounded by $r_e$ and $r_c$, then considering no contribution from the boundary terms for smooth configurations could lead to the identity
\begin{equation}
\int_V d^4x\sqrt{-g}\mu^2_{\text{eff}}\phi=0
\end{equation}
which implies that $\phi$ must change sign in the integration interval $r_e<r<r_c$ for possible non-trivial scalar fields. Thus, the node $k$ of the scalar field have to be $k\geq1$, such that for dS case, the lowest state of the scalar field from which a scalarized black hole may emerge is the first excited state with node $k=1$, independent of the mass of scalar field and the modes with different $\ell$.

Then here we shall focus on $\ell=0$ mode for convenience and mainly study the effect of the scalar mass on the scalarized black hole.

\begin{figure}[H]
 \centering
  \includegraphics[width=5cm]{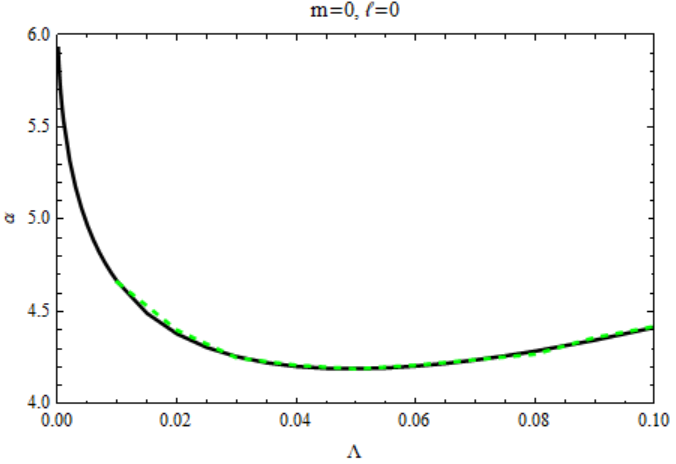}\hspace{0.5cm }
  \includegraphics[width=5cm]{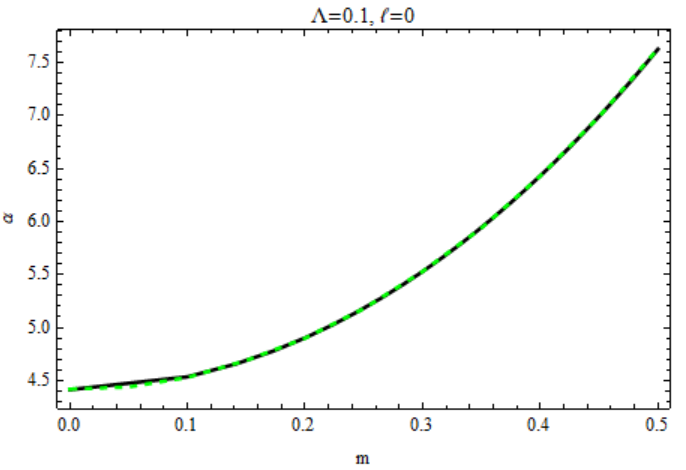}\hspace{0.5cm }
  \includegraphics[width=5cm]{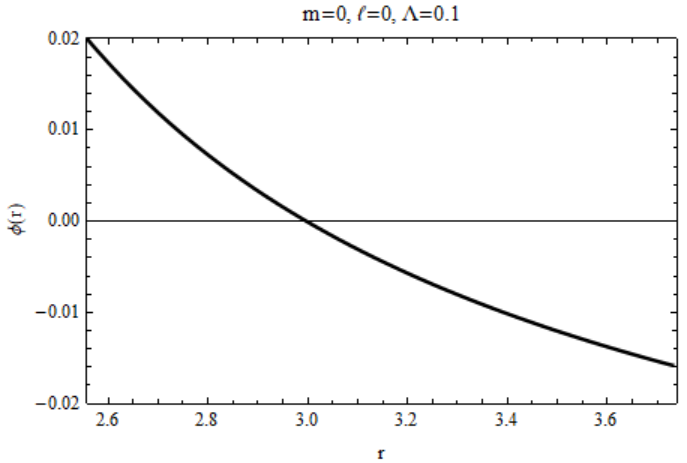}
  \caption{The bifurcation points against $\Lambda$ (left panel) and scalar mass (middle panel) at which it could emerge a scalarized black hole from zero mode with $k=1$. The right panel shows a typical profile of such scalar cloud. We have set $M=1$.}
 \label{fig:cloud-k1}
\end{figure}

We firstly reproduce the results of  \cite{Brihaye:2019gla} on the bifurcation point $\alpha$ against $\Lambda$ for $\ell=0$ massless scalar clouds with node $k=1$, see the left panel of FIG. \ref{fig:cloud-k1}. In the middle panel, we show the bifurcation point  for different scalar mass with fixed $\Lambda=0.1$, which shows that for large $m$, the ratio $\alpha/M^2$ could increase for the bifurcation. Again the green dashed lines are the unstable/stable border  from the analysis of QNM frequency with overtone $n=2$, which is consistent with the bifurcation points as expected. A typical profile of such scalar cloud is shown in right panel.

\begin{figure}[H]
 \centering
  \includegraphics[width=4.5cm]{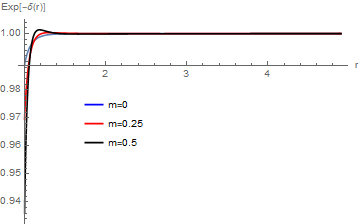}\hspace{0.5cm }
  \includegraphics[width=4.5cm]{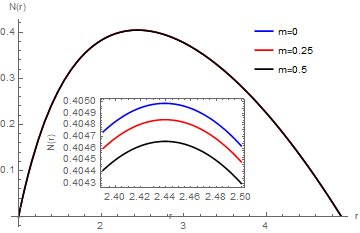}\hspace{0.5cm }
    \includegraphics[width=4.5cm]{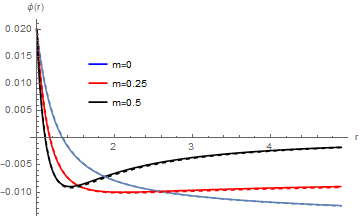}
	  \caption{Profile functions for  typical scalarized solutions with different scalar masses. Here we have set $\Lambda=0.1$ and $\phi(r_e)=0.02$. In the right plot, the solid curves denote the profile for
the scalar field after scalarization while the dashed curves denote the corresponding scalar clouds.}
 \label{fig:hairySolution}
\end{figure}

We proceed to construct the scalarized black hole solution and we will fix the profiles
between the black hole horizon and cosmological horizon. So we consider the scalar field as $\phi=\phi(r)$\footnote{
To construct the scalarized solutions bifurcating from higher modes, one could set $\phi=\sum_{\ell\mathfrak{m}}\phi_{\ell\mathfrak{m}} (r) Y_{\ell\mathfrak{m}} (\theta,\psi)$, and the next steps are straightforward.} and take the ansatz of metric as
\begin{equation}
ds^2=-e^{-2\delta(r)}N(r)dt^2+N^{-1}(r)dr^2+r^2(d\theta^2+\sin^2\theta d\varphi^2),
\end{equation}
where $N(r)$ can be parameterized as
\begin{equation}
N(r)=1-\frac{2\mathcal{M}(r)}{r}-\frac{\Lambda r^2}{3}.
\end{equation}
Then, recalling the equations of motion \eqref{eq:eom1} and \eqref{eq:eom2}, we have the $tt$ and $rr$ components of Einstein equation as
\begin{eqnarray}
m^2 r^2 \varphi ^2-16 \alpha  N^2 \left(\varphi  \varphi ''+\varphi '^2\right)+8 \alpha  \varphi  N' \varphi'+2 r N'+2 \Lambda  r^2-2\nonumber\\
+2 N \left(\varphi  \left(8 \alpha  \varphi ''-12 \alpha  N' \varphi '\right)+\left(8 \alpha +r^2\right)
   \varphi '^2+1\right)=0,\\
   m^2 r^2 \varphi ^2+48 \alpha  N^2 \varphi  \varphi ' \delta '+8 \alpha  \varphi  N' \varphi '+2 r N'+2 \Lambda  r^2-2 \nonumber\\
   -2 N \left(12
   \alpha  \varphi  N' \varphi '+r^2 \varphi '^2+2 \delta ' \left(4 \alpha  \varphi  \varphi
   '+r\right)-1\right)=0,
\end{eqnarray}
and the scalar equation
\begin{eqnarray}
&&\varphi ''+\frac{ \left(rN'+N \left(2-r \delta '\right)\right)}{Nr}\varphi '\nonumber\\
&&+\frac{r^2 \left(-2m^2\phi\right)+8\alpha\phi \left((N-1) \left(N''+2 N \left(\delta '^2-\delta ''\right)\right)+(3-5 N) N' \delta '+N'^2\right)}{4 N r^2}=0.
\end{eqnarray}
To simplify the equations in numeric, we follow the skills in   \cite{Brihaye:2019gla} and  combine the above equations into two first order equations for the metric functions  and a second order equation for scalar field as
\begin{eqnarray}
\delta'&&=F_1(N,\phi,\phi'),\label{eq:backEOM1}\\
N'&&=F_2(N,\phi,\phi'),\label{eq:backEOM2}\\
\phi''&&=F_3(N,\phi,\phi')\label{eq:backEOM3}
\end{eqnarray}
where the formulas of $F_i$ are complex and we will not present here. Then we will numerically integrate the equation groups \eqref{eq:backEOM2}-\eqref{eq:backEOM3} between the event horizon and cosmological horizon. To do so, we should analyze  the approximate form of the solutions at the boundary of our domain of integration. We first solve \eqref{eq:backEOM2} and \eqref{eq:backEOM3} by requiring that $N(r)$ vanishes both at $r_e$ and $r_c$,  and imposing that the scalar field smoothly goes between $r_e$ and $r_c$. Then we substitute the solutions of $N(r)$ and $\phi(r)$ into \eqref{eq:backEOM1} by imposing $\delta$ vanishing at cosmological horizon. With this process, we can numerically construct a branch of scalarized black holes at the bifurcation points from the zero modes. The typical solutions for different scalar masses are shown in FIG. \ref{fig:hairySolution}. We see that the scalar mass has slight effect on the profile of $N(r)$ but the effect on the profile of scalar field is significant. Moreover, from the right panel, we can see the difference between the profile for
the scalar field for the scalarized solution  (solid curves) and  the corresponding scalar clouds (dashed curves).

\begin{figure}[H]
 \centering
  \includegraphics[width=5cm]{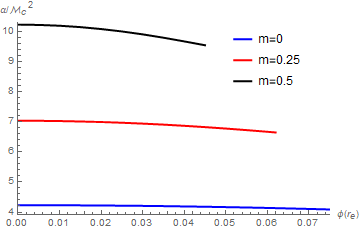}\hspace{0.5cm }
	  \caption{The  `mass' $\mathcal{M}_c=\mathcal{M}(r_c)$  for the scalarized black hole  as a function of  $\phi(r_e)$ for different scalar masses.}
 \label{fig:ph-Mc}
\end{figure}

In the numeric, similar to $\Lambda=0$ case \cite{Silva:2017uqg,Doneva:2017bvd,Antoniou:2017acq} and dS case with massless scalar field \cite{Brihaye:2019gla}, there exists critical configurations that $\phi'(r_e)$ becomes imaginary such that the numerical iterations fail to converge, which is reflected in the existence of critical value of  $\phi(r_e)$.
 Then with fixed $\Lambda=0.1$ we show the  `mass' $\mathcal{M}_c=\mathcal{M}(r_c)$  for the scalarized black hole  as a function of the $\phi(r_e)$ in FIG. \ref{fig:ph-Mc}  where the effect of the scalar mass is explicit.  Here $\phi(r_e)\to 0$ corresponds to the SdS limit while the rightmost dot represents the aforementioned critical configuration where the branches stop to exist.

\section{Conclusion and discussion }
In this paper, we firstly studied the dynamics of the massive scalar field perturbation on Schwarzschild de-Sitter black hole in a special EsGB theory.
In both frequency and time domains, we analyzed the (in)stability of scalar-free dS black hole from the fundamental QNM frequency. To make sure the precision, we first repeat the results in Einstein theory without the GB coupling. In frequency domain, for various $\ell$ perturbation modes, there exists a critical GB coupling $(\alpha_c)$ at which the imaginal part of QNM frequency, $\omega_I$, is zero. When $\alpha<\alpha_c$, $\omega_I$ is negative indicating that the system is stable; while when $\alpha>\alpha_c$, $\omega_I$ turns to be positive implying that the system would undergo a tachyonic instability. Our calculation showed that larger angular momentum, cosmological constant and the mass of scalar field correspond to larger $\alpha_c$, which means that in those cases, the SdS black hole is more difficult to be destabilized. The physical reason is that larger $\alpha$ always give more deeper negative potential well which triggers the tachyonic
instability and finally destroy the SdS black hole, while larger $\ell$, $\Lambda$ and $m$ provide positive values into the effective potential and enhance the negative potential well.  In time domain, for the case with $\alpha<\alpha_c$, the scalar field perturbation would finally decay as time evolves while it would grow up for the case with $\alpha>\alpha_c$. The growth of the perturbation could trigger the scalar-free black hole unstable and a scalarized hairy solution may emerge.
y scanning the model parameters, we figured out the unstable/stable region in the $(\Lambda,\alpha)$-plane and also in the $(m,\alpha)$-plane for various perturbation modes. It is noted that though the scalarization of SdS black hole with backreaction has been investigated  in \cite{Bakopoulos:2018nui,Brihaye:2019gla}, there are still many open issues as their authors addressed.  Here we analyzed the (in)stability in probe limit, but our findings could be helpful to further understand the black hole scalarization in dS spacetime. For example, it is only possible to construct the scalarized hairy solution for the parameters in the shading region in FIG. \ref{fig:lambda-alpha} and FIG. \ref{fig:m-alpha}.

Then we investigated the bifurcation points at which the SdS black hole supports spherical scalar clouds. We found that the bifurcation points from the lowest sate of scalar field with node $k=0$ match well with the border of unstable/stable region obtained in the fundamental frequency for $\ell=0,1,2,3$ modes massless/massive scalar perturbation. And for $\ell=0$, the bifurcation points from the scalar field with node $k=1$ match well with the border of unstable/stable region obtained in the frequency with overtone $n=2$ massless/massive scalar perturbation.
Finally, by including the backreaction, we constructed  the scalarized hairy black hole solutions for different scalar masses. We also showed how  the scalar cloud evolves into the hairy solution through the backreaction by comparing its scalar field with the corresponding scalar clouds. It is worthwhile to mention that allthrough this work, we fix our study in the region between the black hole horizon and cosmological horizon, two boundaries for dS black hole which are mostly concerned.

As a simple way to constraint the model parameters of possible spontaneous scalarization, it is interesting to extended our dynamical analysis  into many cases. A straightforward case is the Schwarzschild or AdS black holes in EsGB theory as well as other modified gravities mentioned in the introduction.  The second case is the dynamic of massive scalar filed perturbed on Kerr black hole in EsGB. Though this direction has been investigated in \cite{Zhang:2020pko}, but there the authors only considered the $\ell=1$ mode of massless scalar field. Since besides the tachyonic instability, the superradiant instability may occur in rotating black hole, so considering the mass of the scalar field and different modes will introduce more physics on the fate of scalar-free background. The next but not the last case is to consider the dynamic of charged black hole before it was scalarized. The scalarization of charged black hole has been firstly proposed in \cite{Herdeiro:2018wub} and soon been generalized in \cite{Myung:2018vug,Fernandes:2019rez,Brihaye:2019kvj,Myung:2019oua,Konoplya:2019goy,Fernandes:2019kmh,Guo:2021zed}, and a complete dynamical analysis on the scalar-free charged black hole still deserves to be present.

\begin{acknowledgments}
We appreciate Xi-Jing Wang for helpful discussion. This work is partly supported by  Natural Science Foundation of China under Grant No. 11775036, Fok Ying Tung Education Foundation under Grant No. 171006 and Natural Science Foundation of Jiangsu Province under Grant No.BK20211601. Guoyang Fu is supported by the Postgraduate Research \& Practice Innovation Program
of Jiangsu Province (KYCX20 2973). Jian-Pin Wu is supported by Top Talent Support
Program from Yangzhou University.
\end{acknowledgments}

\bibliography{Instability-SdS202207}

\end{document}